\documentclass[%
 reprint,
 amsmath,amssymb,
 aps,
 prl,
]{revtex4-2}
\usepackage[utf8]{inputenc}
\usepackage[english]{babel}
\usepackage[T1]{fontenc}
\usepackage{amssymb}
\usepackage{amsmath}
\usepackage{amsfonts} 
\usepackage{amsthm}
\usepackage{hyperref}
\usepackage{float}
\usepackage{enumitem}
\usepackage{bbold}
\usepackage{natbib}
\usepackage{orcidlink}
\newtheorem{theorem}{Theorem}
\usepackage[normalem]{ulem}

\newtheorem{conjecture}{Conjecture}

\begin{document}

\title{Tight bound for the total time in digital-analog quantum computation}
\author{Mikel \surname{Garcia de Andoin}\,\orcidlink{0000-0002-5009-7109}}\email{mikel.garciadeandoin@ehu.eus}\affiliation{Department of Physical Chemistry, University of the Basque Country UPV/EHU, Apartado 644, 48940 Leioa, Spain}\affiliation{EHU Quantum Center, University of the Basque Country UPV/EHU, Barrio Sarriena s/n, 48940 Leioa, Spain}\affiliation{TECNALIA, Basque Research and Technology Alliance (BRTA), Astondo Bidea Ed. 700, 48160 Derio, Spain}
\author{Mikel Sanz\,\orcidlink{0000-0003-1615-9035}}\affiliation{Department of Physical Chemistry, University of the Basque Country UPV/EHU, Apartado 644, 48940 Leioa, Spain}\affiliation{EHU Quantum Center, University of the Basque Country UPV/EHU, Barrio Sarriena s/n, 48940 Leioa, Spain}\affiliation{IKERBASQUE, Basque Foundation for Science, Plaza Euskadi 5, 48009 Bilbao, Spain}\affiliation{Basque Center for Applied Mathematics (BCAM), Alameda Mazarredo 14, 48009 Bilbao, Spain}

\begin{abstract}
    Digital-analog quantum computing (DAQC) is a universal computational paradigm that combines the evolution under an entangling Hamiltonian with the application of single-qubit gates. Since any unitary operation can be decomposed into a sequence of evolutions generated by two-body Hamiltonians, DAQC is inherently well-suited for realizing such operations. Suboptimal upper bounds for the total time required to perform these evolutions have been previously proposed. Here, we improve these limits by providing a tight bound for this crucial parameter, which shows a linear dependence with the number of couplings. This result enables a precise estimation of the time resources needed for quantum simulations and quantum algorithms implemented within the DAQC framework, facilitating a rigorous comparison with other approaches.
\end{abstract}

\maketitle

Any quantum computation can be realized by means of the evolution under a Hamiltonian $H_\text{P}$ for a given time $t\geq0$, $e^{-i t H_P}$. Thus, for a quantum system to be able to perform universal quantum computations, it requires to simulate the evolution under an arbitrary Hamiltonian~\cite{Georgescu2014QuantumSimulation}. One paradigm that allows for this is digital-analog quantum computing, which proposes the use of a universal set of quantum gates from which we can construct any unitary evolution~\cite{Deutsch1995}. Another paradigm corresponds to analog quantum computing, which proposes designing a controllable quantum system that mimics the dynamics of the target evolution \cite{Manousakis2002, Porras2004}. 

There is a possibility of merging both paradigms, as for achieving universality it is sufficient to implement a two-body entangling Hamiltonian and arbitrary single qubit gates~\cite{Dodd2002UnivQC}. This idea was later refined as the digital-analog quantum computing paradigm (DAQC) \cite{Adrian2020DAQC, Galicia2020EnhancedConnect}, which combines the application of single qubit gates with the free evolution under the natural interaction Hamiltonian of the system. In this manner, DAQC benefits from both the flexibility of digital quantum computing and the robustness of analog quantum computing. DAQC has shown resilience against typical error sources \cite{garcia2022noise,canelles2023benchmarking}. As a universal quantum computing paradigm, it allows for the implementation of any quantum algorithm or procedure, such as the quantum Fourier transform \cite{Ana2020}, the Harrow-Hassidim-Lloyd algorithm \cite{Ana2023}, counterdiabatic protocols \cite{pranav2024lyapunov} or quantum convolutional neural networks \cite{Anton2024QCNN}. Several experimental proposals have been made for various platforms \cite{Lamata2018,Tasio2021DAQCcrossResonance,Llenar2024QGARydberg,Kumar2025CDTrappedIons}, and it has already been implemented in superconducting circuits \cite{Babukhin2020, JianWeiPan2023} and trapped ions setups \cite{lu2023trappedions}.

One of the key points that differentiates DAQC from the other two paradigms is the compilation process. In DAQC, the target quantum process can be implemented as one or various evolutions under a two-body interaction Hamiltonian, $U=e^{-iTH}$. By means of the usual Trotter decomposition \cite{Suzuki1976Trotter}, we can isolate the single body terms from this evolution and implement them by using single qubit gates. Then, the evolution that needs to be simulated corresponds to the evolution under a Hamiltonian with two-body interactions,
\begin{equation}
    H_\text{P}=\sum_{i<j}\sum_{\mu,\nu\in\{x,y,z\}} {h_\text{P}}_{ij}^{\mu\nu}\sigma_i^\mu\sigma_j^\nu.
\end{equation}
For this, we have access to the natural evolution of the system, which will have the form 
\begin{equation}
    H_\text{S}=\sum_{i<j}\sum_{\mu,\nu\in\{x,y,z\}} {h_\text{S}}_{ij}^{\mu\nu}\sigma_i^\mu\sigma_j^\nu.
\end{equation}
A key property that we employ to generate DAQC schedules is the fact that if we sandwich an analog block, $e^{-itH_\text{S}}$, with Pauli gates, we get a change of sign in the effective Hamiltonian. As an example, if we sandwich the analog block $e^{-it\sigma_i^z\sigma_k^z}$ with a Pauli $x$ gate applied to one of the two qubits, we can flip the effective sign of the Hamiltonian, $\sigma_i^x e^{-it\sigma_i^z\sigma_k^z}\sigma_i^x=e^{it\sigma_i^z\sigma_k^z}$.
Using this, and assuming a first order Trotter error, a DAQC circuit can be written as
\begin{equation}
    \prod_k V_k^\dagger e^{-it_kH_\text{S}} V_k=\prod_k e^{-i t_k H_\text{S}^{(k)}}\approx e^{-i\sum_kt_kH_\text{S}^{(k)}},
\end{equation}
where $V_k$ is the unitary evolution corresponding to the single qubit gates, and $H_\text{S}^{(k)}$ is the effective system Hamiltonian during the $k$th digital-analog block. 

The last step of the compilation process is to find a set of analog block times $t_k$ that solves the equation 
\begin{equation}
    e^{-iTH_\text{P}}\approx e^{-i\sum_kt_kH_\text{S}^{(k)}}\ \longrightarrow\  T\,H_\text{P}=\sum_kt_kH_\text{S}^{(k)}.
\end{equation}
We can further refine this expression by vectorizing the Hamiltonians. Instead of working in the Hamiltonian's Hilbert space, we can solve the equation by equating the couplings. For this, we describe the Hamiltonian as a column vector of its couplings, $H_\text{P}\rightarrow h_\text{P}=\{{h_\text{P}}_{ij}^{\mu\nu}\}_{ij\mu\nu}$ and equivalently for $H_\text{S}$. Additionally, we can keep the information about the effective signs of the couplings in a matrix $M$, where its columns represent each of the analog blocks and its rows represent each of the couplings. This way, we can write the equation as a simple linear system of equations,
\begin{equation}\label{eq:DAQCsystem}
    T\,h_\text{P}=(M\odot t)\cdot h_\text{S}\ \longrightarrow\ M\,t=T\,h_\text{P}\oslash h_\text{S},
\end{equation}
where $\odot$ and $\oslash$ are the Hadamard or element-wise vector multiplication and division respectively. If a coupling is missing from the system, but the corresponding coupling in the problem is non-zero, ${h_\text{P}}_{ij}^{\mu\nu}\neq{h_\text{S}}_{ij}^{\mu\nu}=0$, the system of equations is not well-defined. Thus, we need to apply a strategy to rearrange the system in order to be solvable, which can consist of swap strategies if a coupling between qubits $i$ and $j$ is missing \cite{Galicia2020EnhancedConnect} or rotations of the Hamiltonian if a $\sigma^\mu\sigma^\nu$ component is missing \cite{MikelAlvaro2024, Bassler2024efficient}. If both couplings are zero, ${h_\text{P}}_{ij}^{\mu\nu}={h_\text{S}}_{ij}^{\mu\nu}=0$, the optimal solution is to remove the coupling from the system of equations \cite{MikelAlvaro2024}. There is also the option of treating the indeterminate form as ${h_\text{P}}_{ij}^{\mu\nu}/{h_\text{S}}_{ij}^{\mu\nu}=0/0=0$, possibly obtaining a suboptimal solution in time \cite{MikelAlatz2025}. For simplicity of notation, we will simplify the notation $b\equiv T\,h_\text{P}\oslash h_\text{S}$ throughout the text. 

The initial proposal for DAQC protocols employed combinations of pairs of $\sigma^x$ gates sandwiching a ZZ-Ising Hamiltonian \cite{Adrian2020DAQC} to attain an arbitrary ZZ-Ising Hamiltonian. A latter proposal, introduced the notion of optimizing the total circuit time by introducing the minimization of $\lVert t\rVert_1$ as the objective of the optimization \cite{Galicia2020EnhancedConnect}, where $\lVert\cdot\rVert_1$ is the vector 1-norm. Throughout the text, we employ the usual $L_p$ norms, $\lVert x\rVert_p=(\sum_{i=1}^n\lvert x_i\rvert^p)^{1/p}$. Afterwards, the original DAQC protocol was extended to arbitrary two-body Hamiltonians \cite{MikelAlvaro2024}. If one includes not only the Pauli basis, but also the single qubit gates from the Clifford group, one might further optimize the DAQC circuits \cite{Bassler2024efficient}. 

An important question is the following: how much time does it take to solve a problem with the optimal DAQC circuit for the worst case? This is equivalent to finding an upper bound for $\lVert t_\text{opt}\rVert_1$. The first result about this was given in Ref.\cite{Bassler2024TimeOptimal}, which proposed an upper bound given by $\lVert t_\text{opt}\rVert_1\leq 2T\lVert h_\text{P}\oslash h_\text{S}\rVert_1$. This bound grows quadratically with the number of qubits in the system, which is far from optimal. In fact, the authors proposed there the conjecture for a tighter bound scaling linearly with the number of qubits,
\begin{conjecture}[Ba{\ss}ler-Heinrich-Kliesch \cite{Bassler2024TimeOptimal}]
    The optimal circuit time for a DAQC protocol with $ZZ$-Ising Hamiltonians is tightly upper bounded by
    \begin{equation}
        \lVert t_\text{opt}\rVert_1\leq T\lVert h_\text{P}\oslash h_\text{S}\rVert_\infty \cdot\begin{cases}
            n, &\text{for odd }n,\\
            n-1, &\text{for even }n.
        \end{cases}
    \end{equation}
\end{conjecture}

Here, we show that this bound is actually not optimal, provide a counter-example, and prove a tight upper bound for the optimal DAQC circuit time: $\lVert t_\text{opt}\rVert_1\leq T\sqrt{3}\lVert h_\text{P}\oslash h_\text{S}\rVert_2$. We achieve this by taking a geometrical perspective based on the properties of convex polytopes. These techniques allow us to construct the explicit form of the problems saturating the bound. Finally, we provide some numerical evidences to support our results.

\textit{Tight upper bound for $t_A$}---
An arbitrary problem for a two-body Hamiltonian is defined by a vector $b\in\mathbb{R}^d$, with $d$ the number of couplings in the problem. Then, the problem of finding a valid DAQC protocol is equivalent to solving the following linear system of equations,
\begin{equation} \label{eq:problem}
    M\,t=b, \text{ with } t\geq0.
\end{equation}
$M \in \mathcal{M}_{d\, d'}(\pm1)$ is a full-rank with $d\leq d'\leq 4^n$. Therefore, the columns of $M$ span $\mathbb{R}^d$, but they are in general not linearly independent. From a geometric point of view, solving Eq.~\eqref{eq:problem} is equivalent to finding the coordinates of the vector $b$ in terms of the aforementioned generating set. In Ref.~\cite{MikelAlvaro2024} it was proven that we can do this for any vector $b$, but that proof leads to a vector of times $t$ such that both the dimension and every element scale exponentially with the system size. Here, we aim for a construction leading to a tighter upper bound for the optimal analog-block time $\lVert t_\text{opt}\rVert_1$.

\begin{theorem}[Upper bound for time in two-body Hamiltonians] \label{thm:Thm1}
    Let us consider the problem of simulating the evolution for a time $T$ under the time independent Hamiltonian $H_\text{P}=\sum_{i,j,\mu,\nu}{h_\text{P}}_{ij}^{\mu\nu}\sigma_i^\mu\sigma_j^\nu$ acting on $n$ qubits. Assume that we employ for that a source Hamiltonian $H_\text{S}=\sum_{i,j,\mu,\nu}{h_\text{S}}_{ij}^{\mu\nu}\sigma_i^\mu\sigma_j^\nu$ also acting on $n$ qubits. For any compatible $H_\text{P}$ and $H_\text{S}$, i.e. ${h_\text{S}}_{ij}^{\mu\nu}=0\Rightarrow{h_\text{P}}_{ij}^{\mu\nu}=0$, for every universal DAQC protocol, there is an implementation whose total analog time $\lVert t_\text{opt}\rVert_1$ is, at most,
    \begin{equation}\label{eq:DAQCbound}
        \lVert t_\text{opt}\rVert_1\leq T\sqrt{3}\lVert h_\text{P}\oslash h_\text{S}\rVert_2,
    \end{equation}
    where $h_\text{P}=\{{h_\text{P}}_{ij}^{\mu\nu}\}_{i,j,\mu,\nu}$  and $h_\text{S}=\{{h_\text{S}}_{ij}^{\mu\nu}\}_{i,j,\mu,\nu}$ are the vector of the couplings of the problem and the source Hamiltonians, respectively. Note that if ${h_\text{P}}_{ij}^{\mu\nu}={h_\text{S}}_{ij}^{\mu\nu}=0$, we remove the corresponding term.
\end{theorem}

The main consequence of this theorem is that the total analog time for any DAQC protocol does not explicitly depend on the number of qubits. However, it does depend on the 2-norm of the vectors of the couplings, which in the worst case, i.e. all-to-all Hamiltonians, scales linearly with the number of qubits. 

Before providing a proof for Theorem \ref{thm:Thm1}, let us provide a geometrical interpretation of the problem. The expression to obtain the DAQC circuit in Eq.~\eqref{eq:DAQCsystem} corresponds to a weighted sum with positive coefficients of the columns of $M$. This way, we can interpret the total analog time $t_\text{A}$ for solving a problem $b=T h_\text{P}\oslash h_\text{S}$ as the sum of its positive coordinates in terms of a generating set which corresponds to the vertices of a convex polytope. With this relation, we construct a demonstration in two parts: first the result is proven for the $ZZ$ Hamiltonian, and afterwards, this result is employed to demonstrate the general case.

\begin{figure}[h]
    \centering
    \includegraphics[width=0.55\linewidth,trim={150pt 66pt 140pt 42pt},clip]{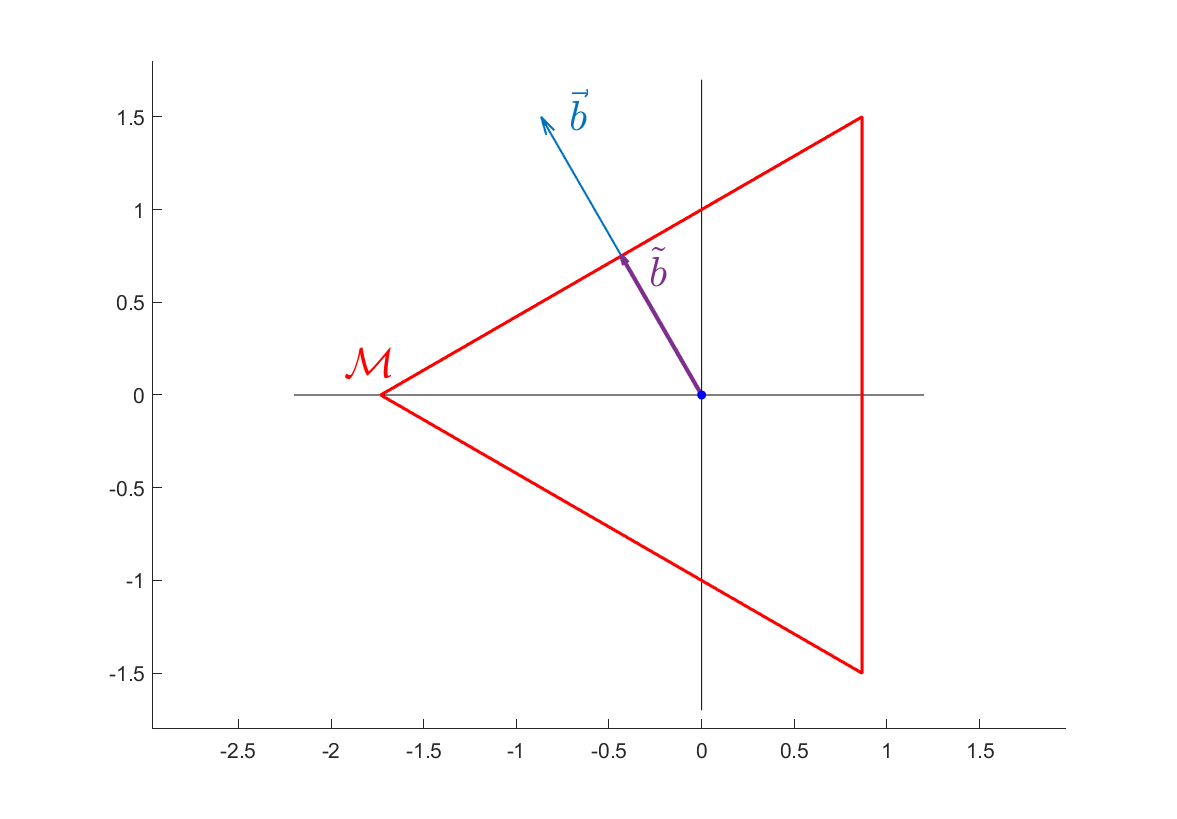}
    \caption{Sketch showing the intuition for building any valid solution for any DAQC problem $\vec{b}$. The problem $\tilde{b}\sim \vec{b}$ can be solved in a total analog block time $\lVert t\rVert_1\leq1$, as it is on the surface of the polytope, $\tilde{b}\in\mathcal{M}$.}
    \label{fig:intuition}
\end{figure}

To build up an intuition for the proof, let us begin by noticing that we can generate a convex polytope $\mathcal{M}\subseteq\mathbb{R}^{n(n-1)/2}$ by taking the columns of the matrix $M$. As proven in Ref. \cite{MikelAlvaro2024}, this polytope contains the origin. This means that we can construct any vector $x\in\mathbb{R}^{n(n-1)/2}$ as a positive sum of the columns of $M$. By definition of a polytope, if $x\in\mathcal{M}$, then we can construct $x$ as a convex combination of the elements in $\mathcal{M}$. If $x$ is outside the polytope generated by the extreme points of the convex hull, then we can normalize $x$ to be in the polytope, calling this new vector $\tilde{x}$. Then,  $\tilde{x}$ can be written as convex combination of the extreme points and rescaled back to $x$ by multiplying it by a constant that depends on the 2-norm, $x=\tilde{x}\lVert x\rVert_2 /\lVert\tilde{x}\rVert_2$. See the sketch in Fig.~\ref{fig:intuition}.

Firstly, let us note that the 2-norm of every point of the polytope with respect to the origin of coordinates is lower bounded by a finite value in a universal DAQC protocol \cite{MikelAlvaro2024}. However, finding this bound is computationally costly. For instance, a naive approach would be to calculate first all the facets of that polytope, and then calculate the minimum distance between each facet and the origin. Unfortunately, this problem is equivalent to the facet enumeration problem, which is NP-hard \cite{Khachiyan2008}. Instead, we will see that the worst possible case corresponds to the smallest non-trivial system sizes (2 and 3 qubits). The intuition comes from two facts. First, the recursive structure of the matrix $M$, employed in Ref.~\cite{MikelAlvaro2024},
\begin{equation}\label{eq:M(n+1)}
    M(n+1) = \begin{pmatrix}
        L(n+1) & L'(n+1)\\
        M(n) & M(n)
    \end{pmatrix},
\end{equation}
where $n$ is the number of qubits, and $L(n)$ and $L'(n)$ correspond to the sign changes of the new couplings added along with the additional qubit. For instance, for the ZZ case, we can assign $L(n+1)$ to the effective signs of the couplings when no gate is applied to the $(n+1)$th qubit, and $L'(n+1)$ to the contrary. In this case, we will get the following relation $L'(n+1)=-L(n+1)$. Second, the number of vertices of the polytope $\mathcal{M}$ grows exponentially with the number of qubits, while the dimension of the space in which it lives grows only quadratically. This gives us the intuition that, as $n$ grows, the polytope resembles more and more a $n(n-1)/2$-dimensional ball. Thus, increasing the number of qubits should not increase the minimum distance from the origin to the surface. Then, from the smallest non-trivial size, we can iteratively increase the number of qubits, and at each step, we will find that the worst possible problem for $n$ qubits corresponds to the worst problem for the smallest $n$ with the additional coefficients zero. The complete proof is provided in the End Matter.

\textit{Numerical calculations}---
In order to numerically confirm the obtained bound, we have run a numerical calculation. For this, we have generated different problem vectors $b$, all normalized with respect to the columns of $M$, $\lVert b\rVert_2=\lVert M_k\rVert_2$. We further set $T=1$ for simplicity. This normalization assures us that the lower bound of $\lVert t_\text{opt}\rVert_1$ is set to 1 for any system size.

\begin{figure}[ht!]
    \centering
    \includegraphics[width=\linewidth,trim={20pt 0pt 50pt 28pt},clip]{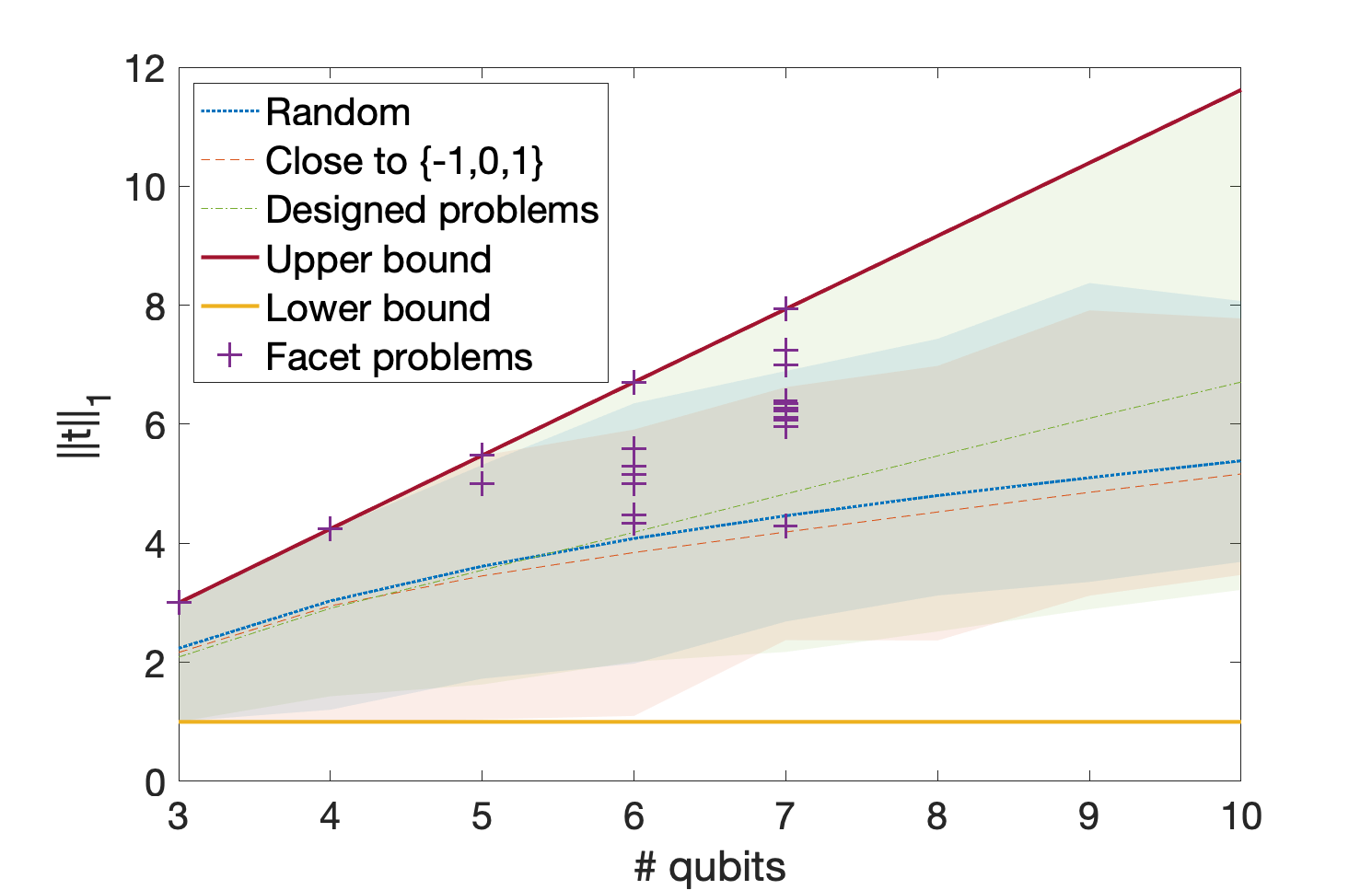}
    \caption{Minimum total analog block times for $ZZ$ Hamiltonians for problems with norm $\lVert b\rVert_2=\sqrt{n(n-1)/2}$. Colored area shows the range between the obtained maximum and minimum total time for the different problem distributions, while the lines represent their mean value.}
    \label{fig:ZZsimulations}
\end{figure}

We start by checking the bound for $ZZ$ Hamiltonians with up to $n=10$ qubits. For this, we have generated $10^6$ random problems uniformly distributed in the surface of a ball of the corresponding dimension. As shown in Fig.~\ref{fig:ZZsimulations}, the generated problems appear to concentrate in the middle between the upper and lower limits, showing a sublinear behavior. As both the upper bound and the lower bound can be reached for problems of the form $b\sim\{\pm1,0\}^{n(n-1)/2}$, we have generated $10^6$ problems in which we uniformly sample all the combinations. To search on a larger sample set, we added a small perturbation uniformly distributed on the $[-0.1,0.1]$ range to those problems. This way of generating problems still concentrates around the same values as obtained with the previous distribution. As an additional note, the lower values obtained tend to be close to the lower bound more, as these corresponds to problems proportional to columns in $M$. In order to approach the upper bound we generate random problems near the axes, but forcing them to have a maximum of 6 nonzero elements. Additionally, we have calculated the solution for the problems that coincide with the closest points of the facets of $\mathcal{M}$ to the origin. However, due to the complexity of calculating the facets, we only made the exact calculation for up to $n=7$ qubits for the $ZZ$ Hamiltonian case. This way, we confirm that the upper bound is reached for some of these problems. These results numerically show that our proposed upper bound is tight.

\begin{figure}[t]
    \centering
    \includegraphics[width=\linewidth,trim={20pt 0pt 50pt 28pt},clip]{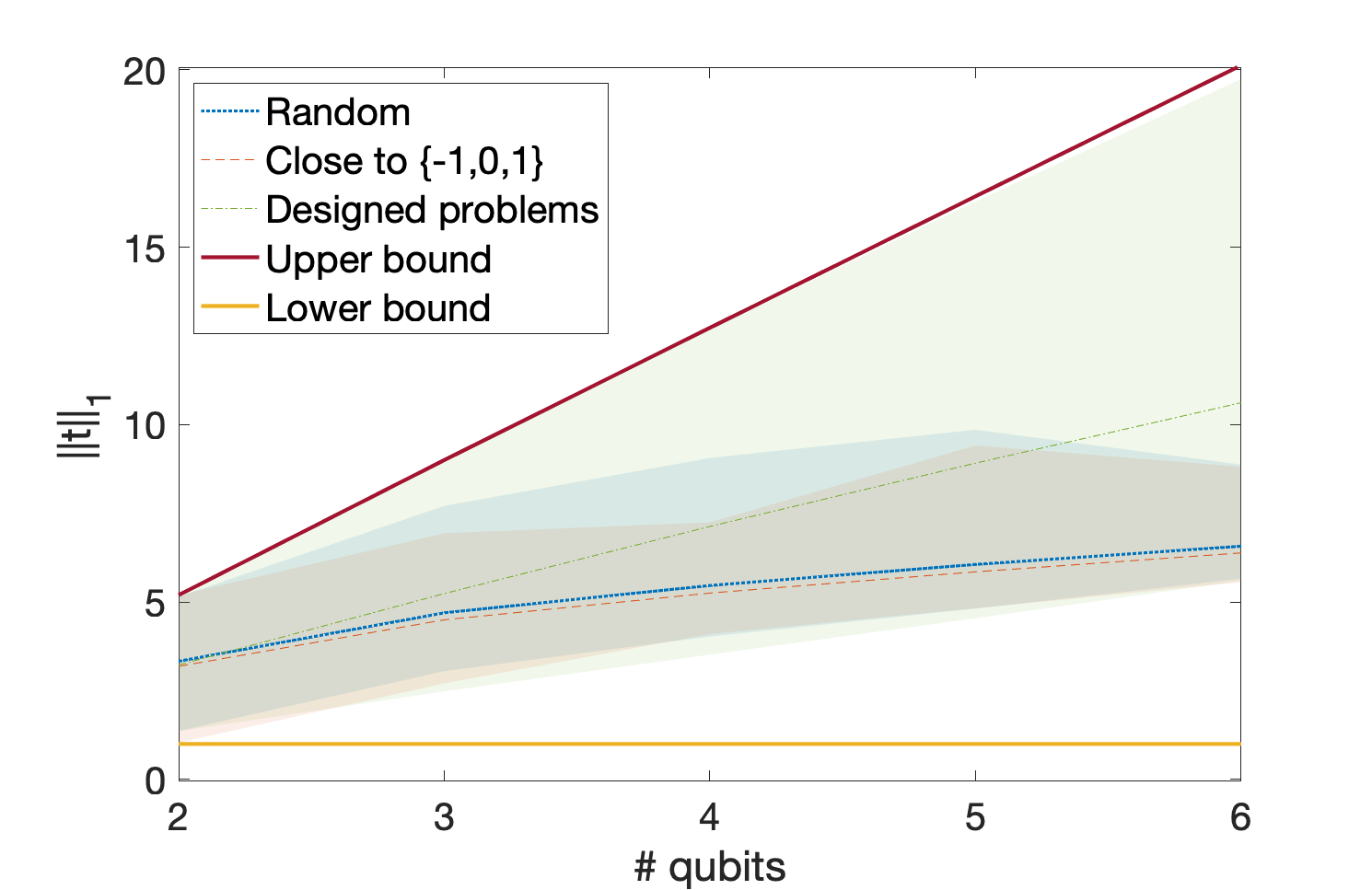}
    \caption{Minimum total analog block times for arbitrary two-body Hamiltonians and problems with norm $\lVert b\rVert_2=\sqrt{9n(n-1)/2}$.}
    \label{fig:arbitrarySimulations}
\end{figure}

We again repeat the same calculation but for arbitrary 2 body Hamiltonians, Fig.~\ref{fig:arbitrarySimulations}. In this case, we do not have access to the information for the facets of $\mathcal{M}$ for $n>4$, so we have only run the calculations for the randomly generated problems. The results in this case are similar to the previous case, further confirming of the tightness of our bound and the sublinearity of uniformly random problems.

In the above, we have obtained a tight upper bound for the total time of an optimal digital-analog schedule. This gives a concise idea of the scaling of DAQC circuits with the number of qubits. As shown, the maximum simulation time for a two-body Hamiltonian increases with the number of couplings. This way, the optimal total time for optimal DAQC circuits lies in the range
\begin{equation}
    T \lVert h_\text{P}\oslash h_\text{S}\rVert_\infty \leq \lVert t_\text{opt}\rVert_1\leq T\sqrt{3} \lVert h_\text{P}\oslash h_\text{S}\rVert_2.
\end{equation}
These bounds are tight, and we can identify the limiting cases. The lower bound is reached in some cases where $\lvert{h_\text{P}}_{ij}^{\mu\nu}/{h_\text{S}}_{ij}^{\mu\nu}\rvert$ is equal for all couplings, leading to a DAQC circuit with a single digital-analog block. The upper bound is reached for some problems in which 3 non-zero couplings are connected with each other, and the rest of the couplings have zero contribution to the problem. The results obtained here are extrapolable to any DAQC protocol which aims at optimizing the total time, both with arbitrary single qubit rotation angles \cite{MikelAlvaro2024} or fixed rotation angles \cite{Adrian2020DAQC, Bassler2024efficient, MikelAlvaro2024}, as the protocols composed of Pauli gates are the worst of the three.  This allows us to estimate the time resources needed to perform a simulation with DAQC, further allowing us to compare it with the time needed for its DQC counterpart. Additionally, the obtained bound may improve error estimations in the circuit. In particular, we can employ different product formulas and their corresponding error bounds to estimate the errors in the DAQC approach \cite{Zhao2022,Layden2022,Zhuk2024,feng2025}. This may help identify the main sources of Trotterization errors and reduce them by using state-of-the-art techniques \cite{faehrmann2022,bagherimehrab2025} or by selecting the appropriate DAQC implementation, improving the practical simulation of longer Hamiltonian dynamics.
In summary, we provide an answer to an open question and confirm the previous intuition regarding a linear dependence on the system size of the upper bound of DAQC circuit times, enabling the further development of this computational paradigm.\\

\textit{Acknowledgements}---
We thank M. Reichert, A. Gomez Tejedor, J. Ferreiro Vélez and D. Aguirre for their useful comments and discussions.
We acknowledge financial support from OpenSuperQ+100 (Grant No. 101113946) of the EU Flagship on Quantum Technologies,
from Project Grant No. PID2024-156808NB-I00 
and Spanish Ramón y Cajal Grant No. RYC-2020-030503-I funded by MICIU/AEI/10.13039/501100011033 and by “ERDF A way of making Europe” and “ERDF Invest in your Future”,
from the Spanish Ministry for Digital Transformation and of Civil Service of the Spanish Government through the QUANTUM ENIA project call-Quantum Spain, and by the EU through the Recovery, Transformation and Resilience Plan--NextGenerationEU within the framework of the Digital Spain 2026 Agenda, 
and from Basque Government through Grant No. IT1470-22,
and the Elkartek project KUBIBIT - kuantikaren berrikuntzarako
ibilbide teknologikoak (ELKARTEK25/79).
MGdA acknowledges support from the UPV/EHU and TECNALIA 2021 PIF contract call, 
and from the Basque Government through the "Plan complementario de comunicación cúantica" (EXP.2022/01341) (A/20220551).

\textit{Data availability}---
The data generated for Fig.\ref{fig:ZZsimulations} and Fig.\ref{fig:arbitrarySimulations} is available upon reasonable request.

\bibliography{main.bib}

@article{Deutsch1995,
author = {Deutsch, David Elieser  and Barenco, Adriano  and Ekert, Artur },
title = {Universality in quantum computation},
journal = {Proc. Roy. Soc. Lond. A Math.},
volume = {449},
number = {1937},
pages = {669-677},
year = {1995},
doi = {10.1098/rspa.1995.0065},
url = {https://doi.org/10.1098/rspa.1995.0065}
}

@article{Lamata2018,
author = {Lucas Lamata and Adrian Parra-Rodriguez and Mikel Sanz and Enrique Solano},
title = {Digital-analog quantum simulations with superconducting circuits},
journal = {Adv. Phys. X},
volume = {3},
number = {1},
pages = {1457981},
year  = {2018},
publisher = {Taylor & Francis},
doi = {10.1080/23746149.2018.1457981}
}

@article{Adrian2020DAQC,
title={Digital-analog quantum computation},
volume={101},
url={http://dx.doi.org/10.1103/PhysRevA.101.022305},
DOI={10.1103/physreva.101.022305},
number={2},
pages={022305},
year={2020},
journal={Phys. Rev. A},
publisher={American Physical Society (APS)},
author={Parra-Rodriguez, Adrian and Lougovski, Pavel and Lamata, Lucas and Solano, Enrique and Sanz, Mikel}
}

@article{garcia2022noise,
author={Garc{\'i}a-Molina, Paula
and Martin, Ana
and Garcia de Andoin, Mikel
and Sanz, Mikel},
title={Mitigating noise in digital and digital--analog quantum computation},
journal={Communications Physics},
year={2024},
month={Oct},
day={06},
volume={7},
number={1},
pages={321},
issn={2399-3650},
doi={10.1038/s42005-024-01812-5},
url={https://doi.org/10.1038/s42005-024-01812-5}
}

@article{Ana2020,
  title = {Digital-analog quantum algorithm for the quantum {F}ourier transform},
  author = {Martin, Ana and Lamata, Lucas and Solano, Enrique and Sanz, Mikel},
  journal = {Phys. Rev. Res.},
  volume = {2},
  issue = {1},
  pages = {013012},
  numpages = {9},
  year = {2020},
  month = {Jan},
  publisher = {American Physical Society},
  doi = {10.1103/PhysRevResearch.2.013012},
  url = {https://link.aps.org/doi/10.1103/PhysRevResearch.2.013012}
}

@article{Ana2023,
  title = {Digital-Analog Co-Design of the {H}arrow-{H}assidim-{L}loyd Algorithm},
  author = {Martin, Ana and Ibarrondo, Ruben and Sanz, Mikel},
  journal = {Phys. Rev. Appl.},
  volume = {19},
  issue = {6},
  pages = {064056},
  numpages = {8},
  year = {2023},
  month = {Jun},
  publisher = {American Physical Society},
  doi = {10.1103/PhysRevApplied.19.064056},
  url = {https://link.aps.org/doi/10.1103/PhysRevApplied.19.064056}
}

@article{Tasio2021DAQCcrossResonance,
title = {Digital-Analog Quantum Simulations Using the Cross-Resonance Effect},
author = {Gonzalez-Raya, Tasio and Asensio-Perea, Rodrigo and Martin, Ana and C\'eleri, Lucas C. and Sanz, Mikel and Lougovski, Pavel and Dumitrescu, Eugene F.},
journal = {PRX Quantum},
volume = {2},
issue = {2},
pages = {020328},
numpages = {20},
year = {2021},
publisher = {American Physical Society},
doi = {10.1103/PRXQuantum.2.020328},
url = {https://link.aps.org/doi/10.1103/PRXQuantum.2.020328}
}

@article{Galicia2020EnhancedConnect,
title = {Enhanced connectivity of quantum hardware with digital-analog control},
author = {Galicia, Asier and Ram{ó}n, Borja and Solano, Enrique and Sanz, Mikel},
journal = {Phys. Rev. Res.},
volume = {2},
issue = {3},
pages = {033103},
numpages = {11},
year = {2020},
publisher = {American Physical Society},
doi = {10.1103/PhysRevResearch.2.033103},
url = {https://link.aps.org/doi/10.1103/PhysRevResearch.2.033103}
}

@article{Dodd2002UnivQC,
title = {Universal quantum computation and simulation using any entangling Hamiltonian and local unitaries},
author = {Dodd, Jennifer L. and Nielsen, Michael A. and Bremner, Michael J. and Thew, Robert T.},
journal = {Phys. Rev. A},
volume = {65},
issue = {4},
pages = {040301(R)},
numpages = {4},
year = {2002},
publisher = {American Physical Society},
doi = {10.1103/PhysRevA.65.040301},
url = {https://link.aps.org/doi/10.1103/PhysRevA.65.040301}
}

@Article{Suzuki1976Trotter,
author={Suzuki, Masuo},
title={Generalized {T}rotter's formula and systematic approximants of exponential operators and inner derivations with applications to many-body problems},
journal={Commun. Math. Phys.},
year={1976},
day={01},
volume={51},
number={2},
pages={183-190},
doi={10.1007/BF01609348},
url={https://doi.org/10.1007/BF01609348}
}

@article{MikelAlvaro2024,
  title = {Digital-analog quantum computation with arbitrary two-body Hamiltonians},
  author = {Garcia-de-Andoin, Mikel and Saiz, \'Alvaro and P\'erez-Fern\'andez, Pedro and Lamata, Lucas and Oregi, Izaskun and Sanz, Mikel},
  journal = {Phys. Rev. Res.},
  volume = {6},
  issue = {1},
  pages = {013280},
  numpages = {14},
  year = {2024},
  month = {Mar},
  publisher = {American Physical Society},
  doi = {10.1103/PhysRevResearch.6.013280},
  url = {https://link.aps.org/doi/10.1103/PhysRevResearch.6.013280}
}

@article{Bassler2024TimeOptimal,
   title={Time-optimal multi-qubit gates: Complexity, efficient heuristic and gate-time bounds},
   volume={8},
   ISSN={2521-327X},
   url={http://dx.doi.org/10.22331/q-2024-03-13-1279},
   DOI={10.22331/q-2024-03-13-1279},
   journal={Quantum},
   publisher={Verein zur Forderung des Open Access Publizierens in den Quantenwissenschaften},
   author={Baßler, Pascal and Heinrich, Markus and Kliesch, Martin},
   year={2024},
   month=mar, 
   pages={1279} 
}

@article{Georgescu2014QuantumSimulation,
  title = {Quantum simulation},
  author = {Georgescu, Iulia and Ashhab, Sahel and Nori, Franco},
  journal = {Rev. Mod. Phys.},
  volume = {86},
  issue = {1},
  pages = {153--185},
  numpages = {33},
  year = {2014},
  month = {Mar},
  publisher = {American Physical Society},
  doi = {10.1103/RevModPhys.86.153},
  url = {https://link.aps.org/doi/10.1103/RevModPhys.86.153}
}

@Article{Manousakis2002,
author={Manousakis, Efstratios},
title={A Quantum-Dot Array as Model for Copper-Oxide Superconductors: A Dedicated Quantum Simulator for the Many-Fermion Problem},
journal={Journal of Low Temperature Physics},
year={2002},
month={Mar},
day={01},
volume={126},
number={5},
pages={1501-1513},
issn={1573-7357},
doi={10.1023/A:1014295416763},
url={https://doi.org/10.1023/A:1014295416763}
}

@article{Porras2004,
  title = {Effective Quantum Spin Systems with Trapped Ions},
  author = {Porras, D. and Cirac, J. I.},
  journal = {Phys. Rev. Lett.},
  volume = {92},
  issue = {20},
  pages = {207901},
  numpages = {4},
  year = {2004},
  month = {May},
  publisher = {American Physical Society},
  doi = {10.1103/PhysRevLett.92.207901},
  url = {https://link.aps.org/doi/10.1103/PhysRevLett.92.207901}
}

@article{Anton2024QCNN,
  title = {Digital-analog quantum convolutional neural networks for image classification},
  author = {Simen, Anton and Flores-Garrigos, Carlos and Hegade, Narendra N. and Montalban, Iraitz and Vives-Gilabert, Yolanda and Michon, Eric and Zhang, Qi and Solano, Enrique and Mart\'{\i}n-Guerrero, Jos\'e D.},
  journal = {Phys. Rev. Res.},
  volume = {6},
  issue = {4},
  pages = {L042060},
  numpages = {8},
  year = {2024},
  month = {Dec},
  publisher = {American Physical Society},
  doi = {10.1103/PhysRevResearch.6.L042060},
  url = {https://link.aps.org/doi/10.1103/PhysRevResearch.6.L042060}
}

@article{JianWeiPan2023,
title = {Quantum neuronal sensing of quantum many-body states on a 61-qubit programmable superconducting processor},
journal = {Science Bulletin},
volume = {68},
number = {9},
pages = {906-912},
year = {2023},
issn = {2095-9273},
doi = {https://doi.org/10.1016/j.scib.2023.04.003},
url = {https://www.sciencedirect.com/science/article/pii/S2095927323002359},
author = {Ming Gong and
          He-Liang Huang and
          Shiyu Wang and
          Chu Guo and
          Shaowei Li and
          Yulin Wu and
          Qingling Zhu and
          Youwei Zhao and
          Shaojun Guo and
          Haoran Qian and
          Yangsen Ye and
          Chen Zha and
          Fusheng Chen and
          Chong Ying and
          others}
}

@article{lu2023trappedions,
  title = {Implementing Arbitrary Ising Models with a Trapped-Ion Quantum Processor},
  author = {Lu, Yao and Chen, Wentao and Zhang, Shuaining and Zhang, Kuan and Zhang, Jialiang and Zhang, Jing-Ning and Kim, Kihwan},
  journal = {Phys. Rev. Lett.},
  volume = {134},
  issue = {5},
  pages = {050602},
  numpages = {7},
  year = {2025},
  month = {Feb},
  publisher = {American Physical Society},
  doi = {10.1103/PhysRevLett.134.050602},
  url = {https://link.aps.org/doi/10.1103/PhysRevLett.134.050602}
}

@article{Babukhin2020,
  title = {Hybrid digital-analog simulation of many-body dynamics with superconducting qubits},
  author = {Babukhin, Danila V. and Zhukov, Andrey A. and Pogosov, Walter V.},
  journal = {Phys. Rev. A},
  volume = {101},
  issue = {5},
  pages = {052337},
  numpages = {14},
  year = {2020},
  month = {May},
  publisher = {American Physical Society},
  doi = {10.1103/PhysRevA.101.052337},
  url = {https://link.aps.org/doi/10.1103/PhysRevA.101.052337}
}

@article{Kumar2025CDTrappedIons,
doi = {10.1088/2058-9565/ad8b64},
url = {https://dx.doi.org/10.1088/2058-9565/ad8b64},
year = {2024},
month = {nov},
publisher = {IOP Publishing},
volume = {10},
number = {1},
pages = {015023},
author = {Kumar, Shubham and Hegade, Narendra N and Henrique de Oliveira, Murilo and Solano, Enrique and Gomez Cadavid, Alejandro and Albarrán-Arriagada, F},
title = {Digital-analog counterdiabatic quantum optimization with trapped ions},
journal = {Quantum Science and Technology}
}

@article{Llenar2024QGARydberg,
  title = {Digital-analog quantum genetic algorithm using Rydberg-atom arrays},
  author = {Llenas, Aleix and Lamata, Lucas},
  journal = {Phys. Rev. A},
  volume = {110},
  issue = {4},
  pages = {042603},
  numpages = {12},
  year = {2024},
  month = {Oct},
  publisher = {American Physical Society},
  doi = {10.1103/PhysRevA.110.042603},
  url = {https://link.aps.org/doi/10.1103/PhysRevA.110.042603}
}

@article{pranav2024lyapunov,
doi = {10.1088/2058-9565/ae7d4e},
url = {https://doi.org/10.1088/2058-9565/ae7d4e},
year = {2026},
month = {jun},
publisher = {IOP Publishing},
volume = {11},
number = {3},
pages = {035028},
author = {Chandarana, Pranav and Paul, Koushik and Ranjan Swain, Kasturi and Chen, Xi and del Campo, Adolfo},
title = {Lyapunov controlled counterdiabatic quantum optimization},
journal = {Quantum Science and Technology}
}

@article{canelles2023benchmarking,
    title={Benchmarking digital--analog quantum computation for the inhomogeneous two-body Ising model},
    author={Canelles, Vicente Pina and Algaba, Manuel G and Heimonen, Hermanni and Papi{\v{c}}, Miha and Ponce, Mario and R{\"o}nkk{\"o}, Jami and Thapa, Manish J and de Vega, In{\'e}s and Auer, Adrian},
    journal={Quantum Science and Technology},
    volume={10},
    number={3},
    pages={035029},
    year={2025},
    publisher={IOP Publishing},
    doi={10.1088/2058-9565/add8b6}
}

@article{Bassler2024efficient,
  title = {General, Efficient, and Robust Hamiltonian Engineering},
  author = {Ba\ss{}ler, P. and Heinrich, M. and Kliesch, M.},
  journal = {PRX Quantum},
  volume = {6},
  issue = {4},
  pages = {040346},
  numpages = {33},
  year = {2025},
  month = {Nov},
  publisher = {American Physical Society},
  doi = {10.1103/9yxv-tdqr},
  url = {https://link.aps.org/doi/10.1103/9yxv-tdqr}
}

@misc{MikelAlatz2025,
author={Garcia de Andoin, Mikel and {Á}lvarez Ahedo, Alatz and Franco Rubio, Adri{á}n and Sanz, Mikel},
title={Impact and mitigation of Hamiltonian characterization errors in digital-analog quantum computation},
year={2025},
eprint={2505.03642},
archivePrefix={arXiv},
primaryClass={quant-ph},
url={https://arxiv.org/abs/2505.03642}
}

@Article{Khachiyan2008,
author={Khachiyan, Leonid
and Boros, Endre
and Borys, Konrad
and Elbassioni, Khaled
and Gurvich, Vladimir},
title={Generating All Vertices of a Polyhedron Is Hard},
journal={Discrete {\&} Computational Geometry},
year={2008},
month={Mar},
day={01},
volume={39},
number={1},
pages={174-190},
issn={1432-0444},
doi={10.1007/s00454-008-9050-5},
url={https://doi.org/10.1007/s00454-008-9050-5}
}

@article{Zhao2022,
  title = {Hamiltonian Simulation with Random Inputs},
  author = {Zhao, Qi and Zhou, You and Shaw, Alexander F. and Li, Tongyang and Childs, Andrew M.},
  journal = {Phys. Rev. Lett.},
  volume = {129},
  issue = {27},
  pages = {270502},
  numpages = {7},
  year = {2022},
  month = {Dec},
  publisher = {American Physical Society},
  doi = {10.1103/PhysRevLett.129.270502},
  url = {https://link.aps.org/doi/10.1103/PhysRevLett.129.270502}
}

@article{Layden2022,
  title = {First-Order Trotter Error from a Second-Order Perspective},
  author = {Layden, David},
  journal = {Phys. Rev. Lett.},
  volume = {128},
  issue = {21},
  pages = {210501},
  numpages = {6},
  year = {2022},
  month = {May},
  publisher = {American Physical Society},
  doi = {10.1103/PhysRevLett.128.210501},
  url = {https://link.aps.org/doi/10.1103/PhysRevLett.128.210501}
}

@article{Zhuk2024,
  title = {Trotter error bounds and dynamic multi-product formulas for Hamiltonian simulation},
  author = {Zhuk, Sergiy and Robertson, Niall F. and Bravyi, Sergey},
  journal = {Phys. Rev. Res.},
  volume = {6},
  issue = {3},
  pages = {033309},
  numpages = {19},
  year = {2024},
  month = {Sep},
  publisher = {American Physical Society},
  doi = {10.1103/PhysRevResearch.6.033309},
  url = {https://link.aps.org/doi/10.1103/PhysRevResearch.6.033309}
}

@misc{feng2025,
      title={Trotterization, Operator Scrambling, and Entanglement}, 
      author={Tianfeng Feng and Yue Cao and Qi Zhao},
      year={2025},
      eprint={2506.23345},
      archivePrefix={arXiv},
      primaryClass={quant-ph},
      url={https://arxiv.org/abs/2506.23345}, 
}

@article{faehrmann2022,
  title={Randomizing multi-product formulas for Hamiltonian simulation},
  author={Faehrmann, Paul K and Steudtner, Mark and Kueng, Richard and Kieferova, Maria and Eisert, Jens},
  journal={Quantum},
  volume={6},
  pages={806},
  year={2022},
  publisher={Verein zur F{\"o}rderung des Open Access Publizierens in den Quantenwissenschaften},
  url={https://doi.org/10.22331/q-2022-09-19-806}
}

@misc{bagherimehrab2025,
      title={Faster Algorithmic Quantum and Classical Simulations by Corrected Product Formulas}, 
      author={Mohsen Bagherimehrab and Luis Mantilla Calderon and Dominic W. Berry and Philipp Schleich and Mohammad Ghazi Vakili and Abdulrahman Aldossary and Jorge A. Campos Gonzalez Angulo and Christoph Gorgulla and Alan Aspuru-Guzik},
      year={2025},
      eprint={2409.08265},
      archivePrefix={arXiv},
      primaryClass={quant-ph},
      url={https://arxiv.org/abs/2409.08265}, 
}

\newpage
\appendix*
\onecolumngrid
\section{End matter}
\twocolumngrid
Before addressing the full proof of Th.~\ref{thm:Thm1}, let us first demonstrate it for the $ZZ$-Hamiltonian. The proof of this case is more intuitive and the generalization is straightforward.

First, let us recall that the columns of the matrix $M$ form a generating set of $\mathbb{R}^{n(n-1)/2}$. This was proven in Ref.~\cite{MikelAlvaro2024}, the columns in $M$ form a convex polytope, denoted by $\mathcal{M}$. The origin is contained in this polytope, but not in the frontier $\text{convHull}(M)\subseteq\mathbb{R}^{n(n-1)/2}$.

By definition, we can write any point of the polytope, $p\in\mathcal{M}$, as $p=\sum_{i}t_i M_i=M\alpha,\ \lVert t\rVert_1=1,\ t_i\geq0\,\forall i$. In particular, we can write any point in the surface of $\mathcal{M}$ using the same expression. Let us denote the points in the surface of $\mathcal{M}$ with $\tilde{b}$.

For any $\tilde{b}$, the solutions $\tilde{t}$ with minimal 1-norm will fulfill $\lVert\tilde{t}\rVert_1=1$. This can be easily proven by contradiction. Assume we can write $\tilde{b}$ as a convex combination $M\tilde{t}=\tilde{b}$ with $\lVert t\rVert_1=\xi<1$. Then, we could generate a new vector $b'=(2-\xi)\tilde{b}$ as the convex combination $Mt'=b'$, with $\lVert t'\rVert_1=(2-\xi)\lVert\tilde{t}\rVert_1=1$. However, this would mean that this new point $b'$ is inside the polytope, contradicting the initial statement that $\tilde{b}$ is in the surface of $\mathcal{M}$.

As the polytope is a compact closed set containing the origin, any line connecting the origin with any other point in the space crosses the convex hull. This allows us to project any point $b$ of the space to the convex hull, $b=\beta \tilde{b},\ \beta\geq0\,\ \forall b\in\mathbb{R}^d$. Then, we can write any vector $b=T h_\text{P}\oslash h_\text{S}$ as $b=\beta \tilde{b}$, and simply rescale the system by multiplying it by $\beta$, $Mt=M(\beta\tilde{t})=\beta\tilde{b}=b$. Using the homogeneity of the norms and the fact that $\tilde{t}$ with $\lVert\tilde{t}\rVert=1$ is the optimal solution for any $\tilde{b}$, we get that the optimal total time to solve $b$ is exactly $\lVert t_\text{opt}\rVert_1=\beta$, with $\beta=\lVert b\rVert_2/\lVert\tilde{b}\rVert_2$. 

As $b$ can be an arbitrary vector in $\mathbb{R}^d$, in order to find the upper bound for $\lVert t\rVert_1$ we need to find the minimum 2-norm of $\tilde{b}$. In the geometrical picture, this problem is equivalent to finding the smallest euclidean distance of the origin to any facet of $\mathcal{M}$, $\min\lVert\tilde{b}\rVert_{2}$. This problem is closely related to the facet-enumeration problem \cite{Khachiyan2008}, which is an NP-hard problem. A simple triangulation algorithm would require $\binom{v}{n(n-1)/2}$ steps to enumerate the facets of a convex hull of $v$ points in $\mathbb{R}^{n(n-1)/2}$. As the number of columns in $M$ is exponential with the number of qubits, $\lvert M\rvert=2^{n-1}$, using a triangulation algorithm is not appropriate for large $n$.

First, we consider the worst-case instance for \(n = 3\), which is the smallest non-trivial system size. We then recursively extend the construction to obtain the corresponding worst-case instance for \(n+1\) qubits. However, the behaviour for \(n < 7\) differs slightly from the general pattern, so we first analyse the range \(3 \leq n \leq 6\) and subsequently treat the case \(n \geq 7\).

For $n=3$, we can just brute-force the solution and calculate the facets of $\mathcal{M}$. For $n=3$, $\mathcal{M}$ is a regular tetrahedron, and the smallest distance of the surface of the convex hull of $M$ is found in 4 directions: $(-1,-1,-1)$, $(-1,1,1)$, $(1,-1,1)$ and $(1,1,-1)$. If we set the problems $b$ proportional to these directions, for example $b=(-\alpha,-\alpha,-\alpha)$, the total optimal time to solve the DAQC protocol is $\lVert t_\text{opt}\rVert_1= \sqrt{3}\lVert b\rVert_2$.

Now, we increase the problem size to $n=4$. We will start from the worst problem in $n=3$, and then we parameterize the additional elements, $b=(\beta_1,\beta_2,\beta_3,-\alpha,-\alpha,-\alpha)$. We can always divide this problem into 4 different problems, $b_1=(\beta_1,0,0,-\alpha_1/3,-\alpha_1/3,-\alpha_1/3)$, $b_2=(0,\beta_2,0,-\alpha_2/3,-\alpha_2/3,-\alpha_2/3)$ and $b_3=(0,0,\beta_1,-\alpha_3/3,-\alpha_3/3,-\alpha_3/3)$. This way, the total time for solving the problem $b$ will be upper bounded by the sum of solving each individual problem, $t_\text{A}(b)\leq\sum_i t_\text{A}(b_i)$. Now, we focus on one of the parameterized problems $b_i$. As seen in Eq.~\eqref{eq:M(n+1)}, the matrix corresponding to these problems can be written as
\begin{equation}\label{eq:partialproblem}
    M'(4)=\begin{pmatrix}
        L(4) & L'(4)\\
        M(3) & M(3)
    \end{pmatrix}.
\end{equation}
As before, we can remove the columns corresponding to the zero elements in the problem vector, and thus
we can take $L(4)$ and $L'(4)$ as row vectors. By construction of the $M(4)$ matrix, we can reorder the columns such that we get $L(4)=\vec{1}$, and by definition, $L'(4)=-\vec{1}$. This way, it is easy to solve each of the problems $b_i$, in which the parameterization that yields the maximum total time corresponds to selecting $\beta_i=0$. This election saturates the lower bound for $t_\text{A}$, and so, it corresponds to the minimum time of the parameterized $b$ problem.

We can repeat this process of iteratively adding a qubit to a system, and parameterizing the new couplings, while keeping the worst $b$ from the previous system size. We parameterize the new couplings and solve every subproblem $b_i$, following the same structure as in the $n=4$ case. In all cases, the structure of each of the partial problems is the same as the one from Eq.~\eqref{eq:M(n+1)}. in order to solve every partial problem, we remove the rows corresponding to the zero entries of the problem vector $b_i$. This leaves us with a matrix $M'$ with linearly dependent columns. By removing these redundant columns, the partial problem we have to solve is identical to the one solved in Eq.~\eqref{eq:partialproblem}. We repeat this process up to problem size $n=7$. Here, there is a change in the topology of the problem, as adding an extra qubit introduces a coupling which is disconnected from the original $n=3$ problem. In this case, when we parameterize all the new couplings, the structure of the problem remains identical when removing redundant columns. For $n>6$ it is even easier to reorder the columns to make the identification $L(n)=\vec{1}$ and $L'(n)=-\vec{1}$, i.e. meaning a vector of $1$'s and $-1$s respectively, as these couplings are disconnected from the original problem. This fact can be verified by explicitly calculating the minimum distance, as in the $n=3$ case.

To complete the proof, we assume that the worst problem corresponds to an $n=3$ all-to-all problem embedded in a larger system, and assume that this holds up to $n$ qubits. We then add a new extra qubit and check that the worst case still corresponds to the problem in $n=3$. For this, we again parameterize the new couplings added to the problem. We notice that we can separate them into two types: couplings that are connected to the original $n=3$ problem and those that are not. This allows us to study this arbitrary size $n$ problem as if it was a $n=6$ problem, as for this size we already have the required topology to represent arbitrary sized problems. As for $n=7$ the worst case is obtained by putting all the weight of the problem into the $n=3$ subproblem, we can extend the upper bound to arbitrary $n$. 

Note that the worst case is not unique. We have based the proof on one of the worst case problems found for $n=3$, but this is not the unique worst case problem. Indeed, we have chosen the vector $(-1,-1,-1)$, but we might have chosen any other three cases, for instance $(-1,1,1)$. Likewise, we can change the indices of the qubits which are connected. This way, we have at least $4\binom{n}{3}$ directions in which we can find the worst cases. However, all these cases lead to the same $t_\text{A}$.

The extension of the proof to arbitrary Hamiltonians can be attained by taking the previous case as an starting point. For arbitrary Hamiltonians, the case of $n=2$ is already non-trivial. Indeed, for this case, we have already problems in a 9-dimensional space which are solved by employing 16 possible combinations of single qubit gates. If we use the same intuition as before, the problems $b$ that yield the worst solutions are the ones located in the center of the facets of the convex hull of the columns of $M$, $\mathcal{M}$. In this particular case, the center of the facets points to the directions with 3 non-zero coordinates. In particular, the non-zero couplings can be found for the couplings $\{(XX,YY,ZZ)$, $(XX,YZ,ZY)$, $(YY,XZ,ZX)$, $(ZZ,XY,YX)$, $(XY,YZ,ZX)$, $(XZ,ZY,YX)\}$. The directions in which these problems yield the worst $\lVert t\rVert_1$ corresponds to the same combinations found for the $ZZ$-Hamiltonian case, i.e. $\{(-1,-1,-1)$, $(-1,1,1)$, $(1,-1,1)$, $(1,1,-1)\}$. An example of one of these problems is the vector $b=(-1,0,0,0,-1,0,0,0,-1)$. 

We follow the same scheme that we employed in the previous section. We start by increasing the number of qubits by 1. Then, starting from the worst case obtained for the previous system size, we parameterize the new couplings. Here, we have 2 options for the worst possible case. One corresponds to the scenario in which we have all the non-zero couplings between the same two qubits, i.e. the case mentioned in the previous paragraph. The other case corresponds to having three couplings connecting the same three qubits, which is a similar situation as in the previous subsection. Note that the proof for the $ZZ$-Hamiltonian obtained in the previous subsection is included in the proof for the arbitrary two-body Hamiltonians. The construction is an extension of the ZZ case. Indeed, for any of these cases, we can construct the matrix $M(n+1)$,
\begin{equation*}
    \small
    M(n+1)=\begin{pmatrix}
        L(n+1) & L'(n+1) & L''(n+1) & L'''(n+1) \\
        M(n) & M(n) & M(n) & M(n)
    \end{pmatrix},
\end{equation*}
where the corresponding $L(n+1)$ matrices corresponds to the signs of the new couplings. As in the $ZZ$ case, when solving the problems $b_i$, we parameterize every new coupling. We can always rearrange the matrix in such a way that we are left with a problem which is identical to the one in Eq.~\eqref{eq:partialproblem}, where we have two $M(n)$ blocks, $L(n+1)=\vec{1}$ and $L'(n+1)=-\vec{1}$. Then, the proof follows the same steps as in the ZZ case.

Again, we assume that the worst case for $n$ qubits corresponds to a problem in which only 3 couplings in the vector $b$ are non-zero. Then, we add a single qubit, and parameterize the new couplings. Then, we set all the new parameters to zero except one. We finish the proof by noticing again that the topology of this problem is the same as the one for $n=6$. By brute-forcing the resolution of this problem, we see that the worst case corresponds to setting the new parameters to zero.
\end{document}